\documentclass[twocolumn,showpacs,preprintnumbers,amsmath,amssymb]{revtex4-2}
\usepackage{graphicx}% Include figure files
\usepackage{dcolumn}% Align table columns on decimal point
\usepackage{bm}% bold math
\usepackage{lineno}
\newcommand{\bef}{\begin{figure}}
\newcommand{\eef}{\end{figure}}

\newcommand{\be}{\begin{equation}}
\newcommand{\ee}{\end{equation}}
\newcommand{\bea}{\begin{eqnarray}}
\newcommand{\eea}{\end{eqnarray}}
\begin{document}

\title{The study of $K^{*0}$ meson production using a multi-phase transport model at RHIC BES energies}

\author{Pranjal Barik$^{1}$, Kadambini Menduli$^{1}$, Aswini Kumar Sahoo$^{2}$, Md. Nasim$^{1}$}
\affiliation{$^{1}$Department of Physical Sciences, Indian Institute of Science Education and Research, Berhampur, India;\\
$^{2}$ Institute of Modern Physics, Chinese Academy of Sciences, Lanzhou, China}

\begin{abstract}
We present the yield, average transverse momentum, and collective flow measurement of $K^{*0}$ resonances in Au+Au collisions at $\sqrt{s_{NN}} = 19.6$, 14.5, and 7.7~GeV using the AMPT model. It is found that, due to hadronic rescattering, the decay daughters of $K^{*0}$ interact with other particles in the medium, causing the yield of reconstructable $K^{*0}$ to be significantly suppressed, especially at low transverse momentum. The model results are compared with recent experimental data from Phase-II of the Beam Energy Scan (BES-II) program at the Relativistic Heavy-Ion Collider. The string-melting version of the AMPT model successfully reproduces the measured $K^{*0}/K$ ratios at all three analysed collision energies.  Interestingly, AMPT calculations that exclude the hadronic phase nevertheless provide a reasonable description of the data, thereby challenging the conventional interpretation that hadronic rescattering is the primary mechanism responsible for suppressing the $K^{*0}/K$ ratio in central heavy-ion collisions. In addition, we find that the $K^{*0}/K$ ratio appears to be largely insensitive to the lifetime of the hadronic phase, whereas the average transverse momentum, $\langle p_{T} \rangle$, of the $K^{*0}$ shows a strong dependence, increasing significantly as the lifetime of the hadronic phase becomes longer.
We further show that the directed flow ($v_1$) of $K^{*0}$ mesons is strongly influenced by hadronic rescattering, whereas the elliptic flow ($v_2$) exhibits only weak sensitivity to hadronic effects. These results establish $K^{*0}$ directed flow as a sensitive probe of the late-stage hadronic medium in heavy-ion collisions. These model calculations therefore provide valuable insight into the underlying physics governing the observed experimental results at RHIC.

\end{abstract}
\maketitle

\section{Introduction}
\label{S:1}
%% into for K*
Short-lived resonances serve as an excellent tool to study the properties of the medium formed in heavy-ion collisions~\cite{Brown_resonance, Markert_resonance, Schaffner, Rapp,star_resonance}. Resonances such as $K^{*0}$ are of particular interest as their lifetime ($\sim$ 4 fm/c) is shorter than the duration of the medium created in heavy-ion collision~\cite{system_life}, and can decay, rescatter, and regenerate within the hadronic phase.\\
The resonance $K^{*0}$ undergoes hadronic decays, specifically $K^{*0}(\overline{K^{*0}})\rightarrow K^{\pm}\pi^{\mp}$, with a branching ratio of 2/3~\cite{pdg}. Following the decay, the daughter kaons and pions have the potential to undergo rescattering interactions with other hadrons in the medium, making it challenging to reconstruct the parent resonance from its decay products. At the same time, the abundant population of kaons and pions within the medium can lead to the creation of the $K^{*0}$ resonance through pseudo-elastic scattering. This phenomenon is referred to as regeneration~\cite{reco_issue_1,reco_issue_2,reco_issue_3,reco_issue_4}.
The interplay between rescattering and regeneration can be observed through the $K^{*0}/K$ ratio. The behaviour of this ratio, specifically its suppression or enhancement, with increasing medium multiplicity, provides insights into the relative significance of hadronic rescattering compared to regeneration processes. A suppression of the $K^{*0}/K$ ratio with increasing medium multiplicity would suggest that hadronic rescattering dominates over regeneration, whereas an enhancement would indicate the prevalence of regeneration over rescattering.
Previous experimental measurements~\cite{star_kstar_2002,star_kstar_2005,star_kstar_2008,star_kstar_2011,phenix_kstar_2014, NA49_kstar_2011, NA61_kstar_2020, NA61_kstar_2021,alice_kstar_2012,alice_kstar_2015,alice_kstar_2017,alice_kstar_2020_1,alice_kstar_2020_2,alice_kstar_2020_3,alice_kstar_2022,kstar_BES} suggest, compared to small collision system (e.g. $p+p$), $K^{*0}/K$  ratios in heavy-ion collisions are smaller, which is often interpreted as the effect of stronger hadronic rescattering in central A+A collisions.

%% into for v1
In addition to the yield, it was predicted that hadronic rescattering may influence the collective flow of short-lived resonances~\cite{Cshen,parida}.  Collective motion among produced particles can be studied by measuring the coefficient in the Fourier expansion of the final-state azimuthal distribution of the produced particle relative to the collision reaction plane~\cite{Ollitrault:1992bk, Poskanzer:1998yz}. The first harmonic coefficient, known as directed flow $v_{1}$, describes a collective sideward motion of emitted particles. The second harmonic coefficient, known as elliptic flow $v_{2}$, describes an elliptical anisotropic motion of emitted particles. Both hydrodynamic and nuclear transport models indicate that $v_{1}$ and  $v_{2}$  are sensitive to details of the expansion during the early stages of the collision fireball, as well as sensitive to the late-stage hadronic interaction~\cite{pheno_kstar_2024}. While the effects of hadronic rescattering on yield and elliptic flow of resonances have been extensively studied~\cite{pheno_kstar_2002,pheno_kstar_2015, pheno_kstar_2016,pheno_kstar_2020,pheno_kstar_2023,pheno_kstar_2025, pheno_kstar_2018, pheno_kstar_2021,pheno_v1_nayak}, their impact on directed flow ($v_1$) has received comparatively little attention~\cite{Cshen, parida}. In this paper, we have studied the production yield and collective flow ($v_{1}$ and $v_{2}$) of $K^{*0}$ resonances in Au+Au collisions using  A Multi-Phase Transport (AMPT) Model~\cite{ampt} at RHIC energies. Model results are compared with the available experimental data. We mainly focus on the results measured by the STAR experiment in the Phase-II of the Beam Energy Scan (BES-II) program~\cite{bes-II}.  \\

This paper is organised as follows. Section II briefly describes the AMPT model employed in this study. Section III presents the AMPT model results for the yields and flows of $K^{*0}$ resonances under different configurations, including varying hadronic-cascade times. Finally, Section IV summarises the main findings and discusses the implications of this work for current experimental measurements in high-energy heavy-ion collisions at RHIC.

\begin{figure}[ht]
\includegraphics[scale=0.4]{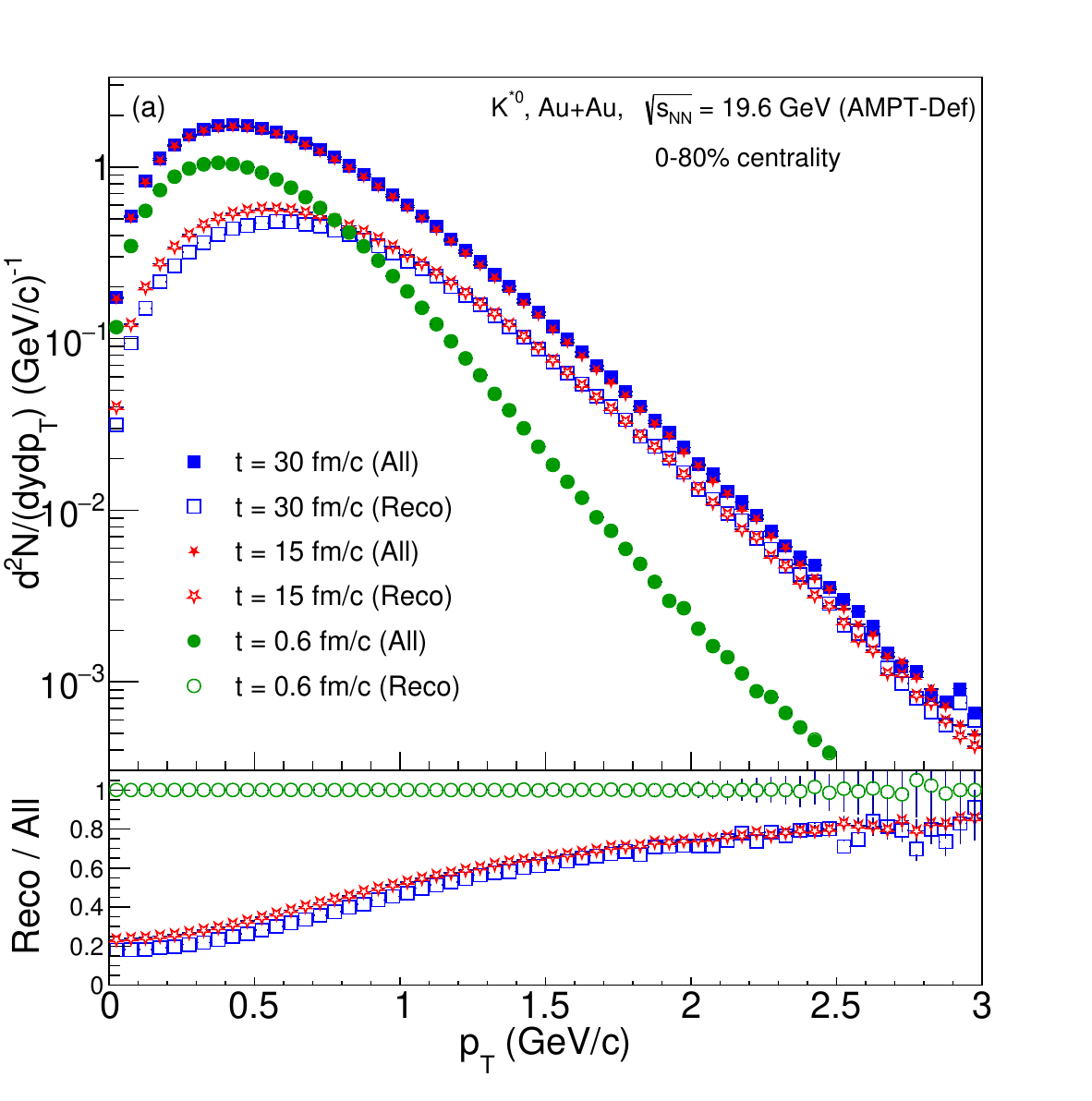}
\includegraphics[scale=0.4]{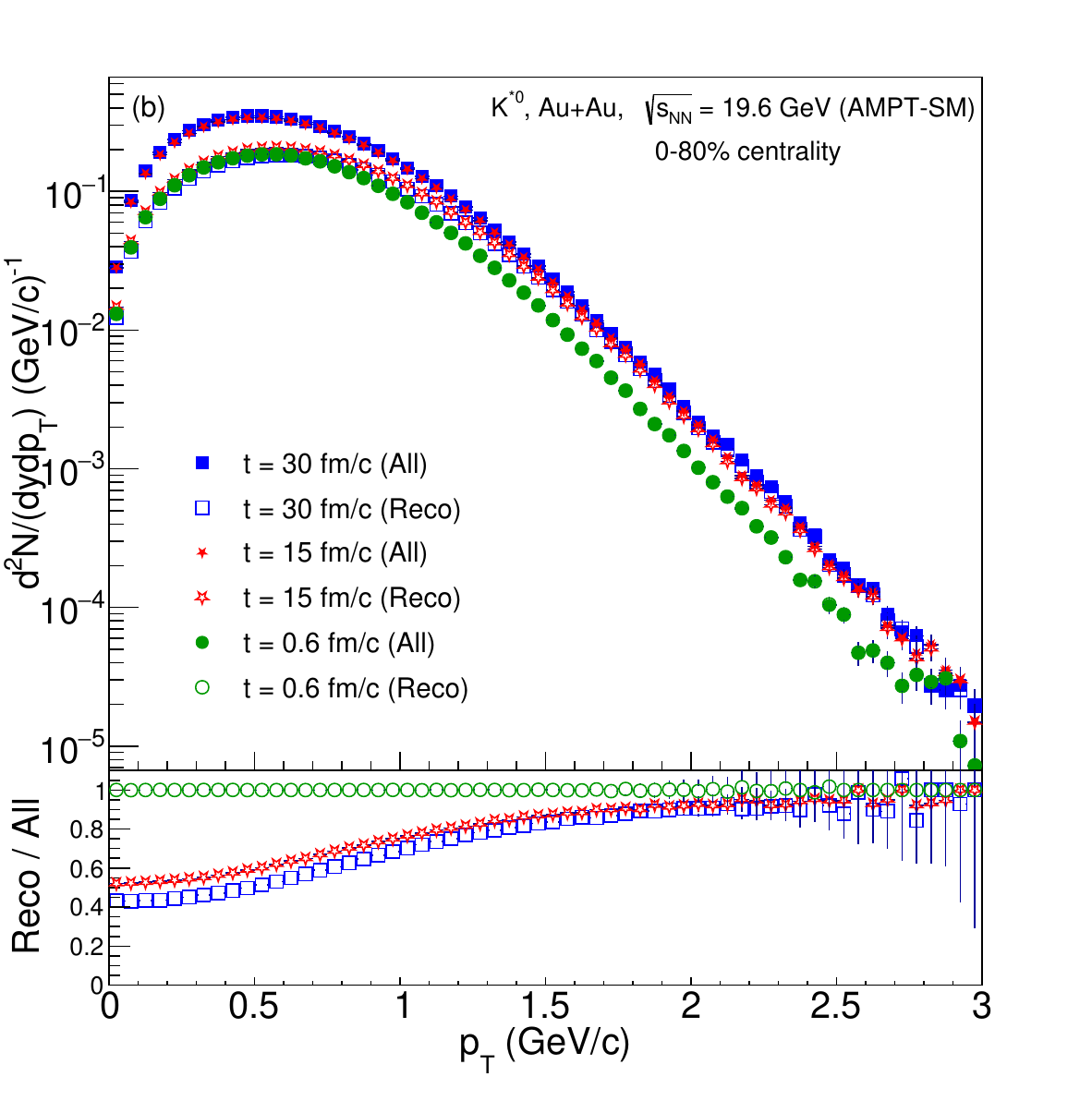}
\caption{ The transverse momentum ($p_{T}$) dependence of the $K^{*0}$ resonance yield in minimum-bias Au+Au collisions at $\sqrt{s_{NN}} = 19.6$~GeV, calculated using the AMPT-SM and AMPT-Def models.  The lower panels show ratios of all produced to reconstructed $K^{*0}$. Calculation is done using the hadronic cascade time $t$ = 0.6, 15, and 30 fm/c.}
\label{fig1}
\end{figure}

\section{Model Description}

The AMPT model used for the calculations presented in this paper has four main stages: the initial conditions, partonic interactions, the conversion from the partonic to the hadronic matter, and hadronic interactions~\cite{ampt}. The initial conditions are obtained from the HIJING model~\cite{hijing}. Scatterings among partons are modelled by Zhang’s parton cascade (ZPC)~\cite{zpc}, which includes only two-body scatterings with cross sections obtained from the pQCD
with screening masses.  The AMPT model can be studied in two configurations: the AMPT default version (labelled as AMPT-Def), in which the minijet partons are made to undergo scattering before they are allowed to fragment into hadrons, and the AMPT string-melting scenario (labelled as AMPT-SM), in which additional scattering occurs among the quarks. The AMPT model with string melting leads to hadron formation using a quark coalescence model. The subsequent hadronic matter interaction is described by a hadronic cascade, which is based on A Relativistic Transport (ART) model~\cite{art}.  Some of the results presented are obtained by varying the termination time of the hadronic cascade from 0.6 fm/c to 30 fm/c to study the effect of the hadronic rescattering on the observables presented. In the AMPT model, a hadronic cascade time of 0.6 fm/c is effectively considered as the absence of hadronic interactions. More detailed discussions regarding the AMPT model can be found in~\cite{ampt}. In this study, approximately one million events for each configuration were generated for Au + Au 0-80\% minimum bias Au+Au collisions at $\sqrt{s_{NN}}$ = 19.6, 14.5, and 7.7~GeV.

\begin{figure*}
\includegraphics[scale=0.75]{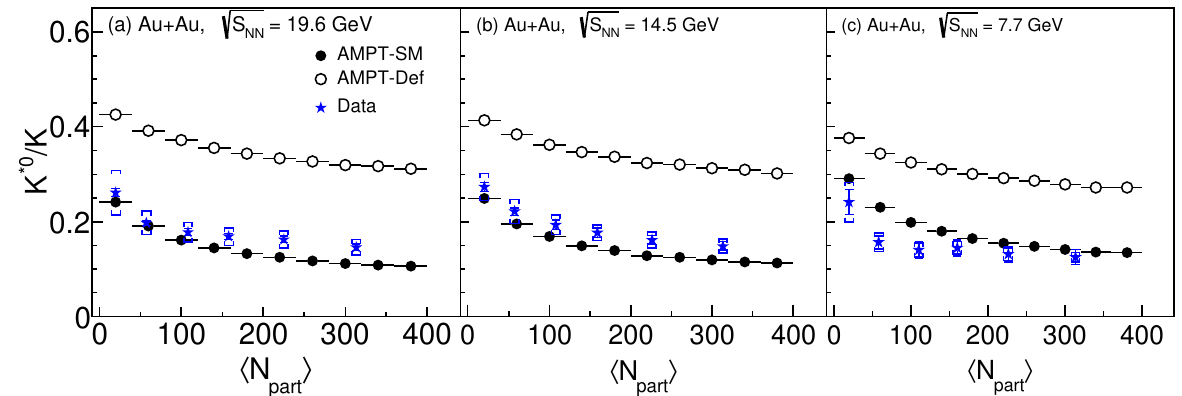}
\caption{The $K^{*0}/K$ ratio [$=(K^{*0}+\overline{K^{*0}})/(K^{+}+K^{-})$] as a function of the number of participating nucleons ($N_{\rm part}$) for Au+Au collisions at $\sqrt{s_{NN}} = 19.6$, 14.5, and 7.7~GeV, calculated using the AMPT model with hadronic cascade time $t$ = 30 fm/c. Experimental data is taken from ~\cite{kstar_bes_II}. }
\label{fig2}
\end{figure*}

\section{Result and DISCUSSION}
%\subsection{Production Yield Calculation}
\subsection{Production Yield }
Figure~\ref{fig1} shows the transverse momentum ($p_{T}$) dependence of the $K^{*0}$ resonance yield in minimum-bias Au+Au collisions at $\sqrt{s_{NN}} = 19.6$~GeV, calculated using the AMPT string-melting (AMPT-SM) and default (AMPT-Def) models. The results are presented for both all produced $K^{*0}$ mesons (denoted as $\rm{All}$) and those reconstructed via the invariant mass method (denoted as $\rm{Reco}$), for different hadronic cascade times $t$.
Here $\rm{All}$ $K^{*0}$ means $K^{*0}$ mesons produced both from coalescence and from the hadronic cascade, regardless of whether they are later destroyed in the hadronic phase. We store both the $K^{*0}$ momentum and the momenta of its daughter particles at each decay position and time. In the invariant mass method, the mass of the parent particle is reconstructed from the measured energies and momenta of the decay daughters using energy–momentum conservation.
In the absence of hadronic interactions ($t = 0.6$~fm/$c$), the number of reconstructed $K^{*0}$ mesons is equal to the number of initially produced ones, as the momenta of the decay daughters remain unmodified, allowing an accurate reconstruction. In contrast, for longer hadronic cascade times ($t = 15$ and 30~fm/$c$), the yield of reconstructable $K^{*0}$ mesons is significantly reduced, particularly at low $p_{T}$. This reduction arises from the rescattering of the decay daughters with other hadrons in the medium, which alters their momenta and prevents successful reconstruction using the invariant mass technique. We also observe that the AMPT-Def model shows a stronger suppression of the $K^{*0}$ yield compared to the AMPT-SM model, with the maximum suppression reaching approximately 80\% in AMPT-Def and about 60\% in AMPT-SM.

Figure~\ref{fig2} shows the $K^{*0}/K$ ratio [$=(K^{*0}+\overline{K^{*0}})/(K^{+}+K^{-})$] as a function of the number of participating nucleons ($N_{\rm part}$) for Au+Au collisions at $\sqrt{s_{NN}} = 19.6$, 14.5, and 7.7~GeV, calculated using the AMPT model. Results from both the default (AMPT-Def) and string-melting (AMPT-SM) versions are presented for a hadronic cascade time of $t = 30$~fm/$c$. Here, $K^{*0}$ yield is calculated using the invariant mass method from the decay daughters. It is observed that the $K^{*0}/K$ ratio is more strongly suppressed in the AMPT-SM model compared to the AMPT-Def model, even though Fig.~\ref{fig1} shows that the loss of reconstructable $K^{*0}$ yield due to rescattering is larger in the AMPT-Def model. This apparent discrepancy arises from differences in the charged kaon production yields predicted by the two models. The AMPT results are also compared with recent experimental measurements from RHIC. We find that the AMPT-SM model successfully describes the experimental data, whereas the AMPT-Def model systematically over-predicts the measured $K^{*0}/K$ ratios. The observed decrease of the $K^{*0}/K$ ratio with increasing collision centrality is commonly interpreted as evidence of enhanced hadronic rescattering in central heavy-ion collisions.

\begin{figure}
\includegraphics[scale=0.40]{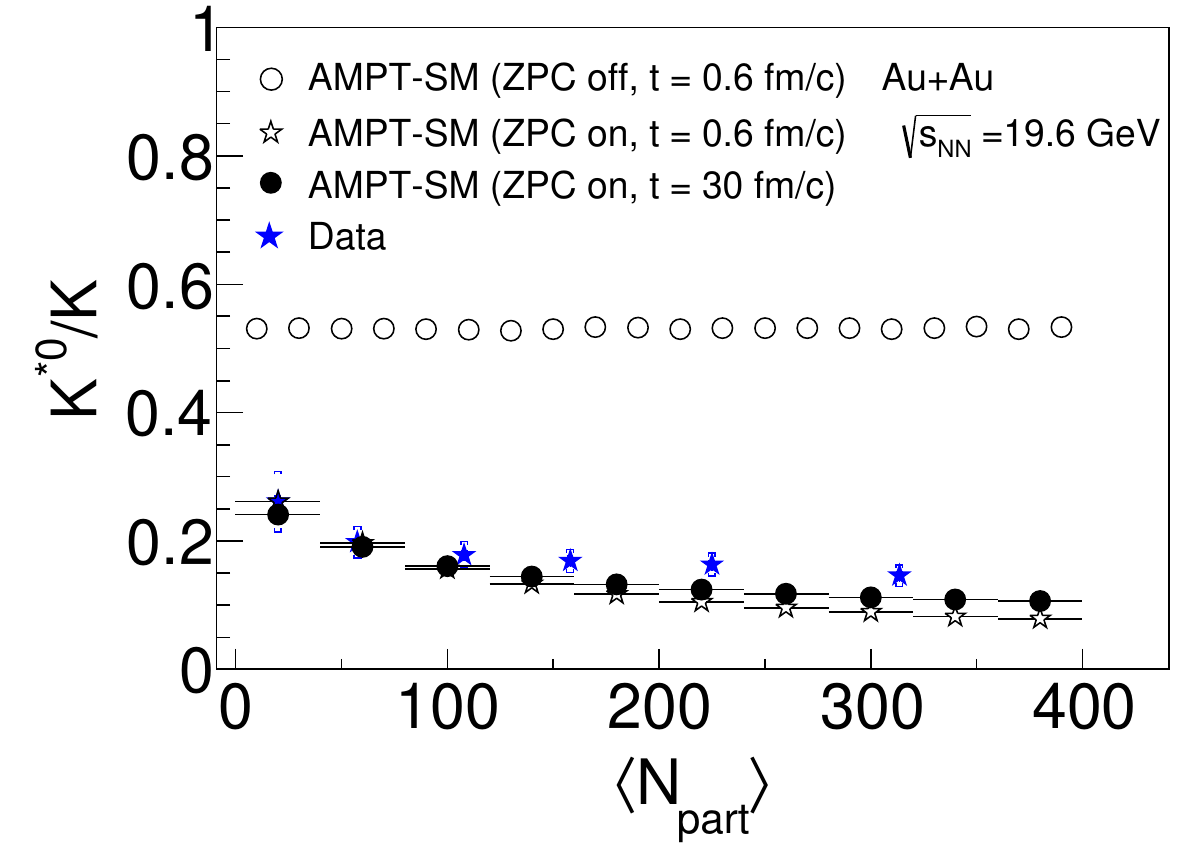}
\caption{ The $K^{*0}/K$ ratio in Au+Au collisions at $\sqrt{s_{NN}} = 19.6$~GeV obtained with the AMPT-SM model for different hadronic cascade times ($t = 0.6$, and 30~fm/$c$) and with and without  Zhang’s parton cascade (ZPC). Experimental data is taken from ~\cite{kstar_bes_II}. }
\label{fig3}
\end{figure}

Figure~\ref{fig3} shows the $K^{*0}/K$ ratio in Au+Au collisions at $\sqrt{s_{NN}} = 19.6$~GeV obtained from the AMPT-SM model for different hadronic cascade times ($t = 0.6$ and 30~fm/$c$), together with the corresponding experimental data. We also present calculations of the $K^{*0}/K$ ratio in the absence of partonic interactions by switching off ZPC in the AMPT model. When both hadronic and partonic interactions are turned off, a flat $K^{*0}/K$ ratio is observed as a function of $N_{\rm part}$. Interestingly, even in the absence of hadronic interactions ($t = 0.6$~fm/$c$), the $K^{*0}/K$ ratio decreases with increasing $N_{\rm part}$ when partonic interactions are present. Similar decresing trend of $K^{*0}/K$ ratio is also observed at 14.5 and 7.7 GeV.  The calculated $K^{*0}/K$ ratio without hadronic interactions is found to be in good agreement with the experimental measurements. These results challenge the conventional interpretation that hadronic rescattering is the dominant mechanism responsible for the suppression of the $K^{*0}/K$ ratio in central heavy-ion collisions. Furthermore, the $K^{*0}/K$ ratio is observed to be nearly independent of the hadronic cascade time. This behaviour arises because regeneration of $K^{*0}$ resonances in the hadronic medium compensates for the loss due to rescattering, even though the yield of reconstructable $K^{*0}$ decreases with increasing $t$.

\begin{figure*}
\includegraphics[scale=0.7]{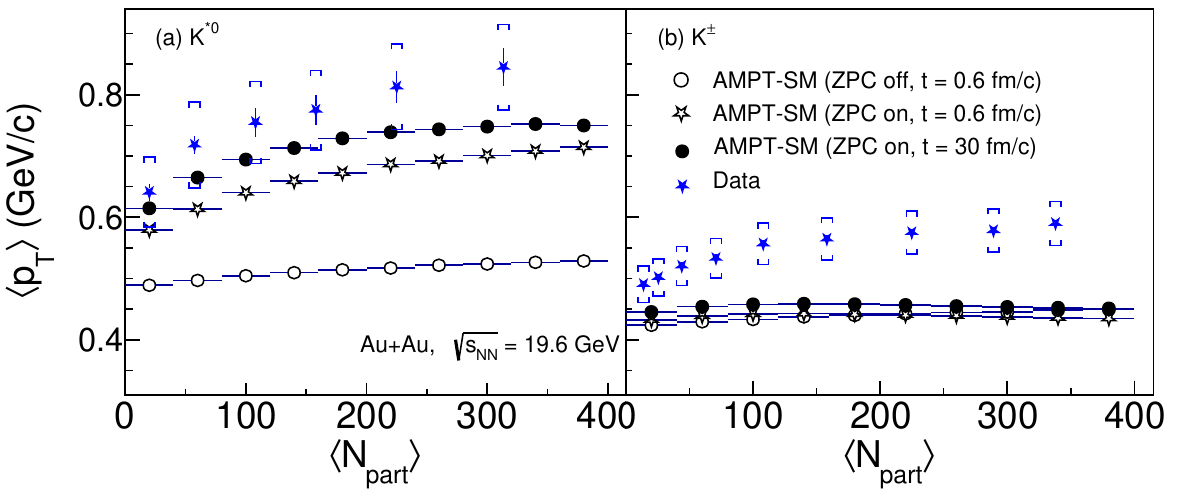}
\caption{ The mean $p_{T}$ $K^{*0}$ and $K^{\pm}$  in Au+Au collisions at $\sqrt{s_{NN}} = 19.6$~GeV obtained with the AMPT-SM model for different hadronic cascade times ($t = 0.6$, and 30~fm/$c$) and with and without  Zhang’s parton cascade (ZPC). Experimental data are taken from ~\cite{kstar_BES,pikp_2017}. }
\label{fig3a}
\end{figure*}

The average transverse momentum ($\langle p_{T}\rangle$) of $K^{*0}$ and $K^{\pm}$ is presented in Fig.~\ref{fig3a}. The results are obtained for Au+Au collisions at $\sqrt{s_{NN}} = 19.6$~GeV using the AMPT-SM model, considering different hadronic cascade durations ($t = 0.6$ and 30~fm/$c$), both with and without partonic interactions. Although the $K^{*0}/K$ ratio shows little dependence on the hadronic cascade time, the $\langle p_{T}\rangle$ of $K^{*0}$ exhibits a stronger sensitivity to the hadronic phase lifetime compared to that of charged kaons. A comparison of the AMPT results with experimental measurements~\cite{kstar_BES,pikp_2017} indicates that the model underestimates the observed values. This discrepancy can be reduced by tuning the initial parameters—particularly those associated with the Lund string fragmentation in the HIJING model—which allows modification of $\langle p_{T}\rangle$ while keeping the particle yields approximately unchanged~\cite{ampt_2014,nasim_tiwari}.    

%  Interestingly, the $K^{*0}/K$ ratio is found to be nearly independent of the hadronic cascade time, despite the fact that the yield of reconstructable $K^{*0}$ resonances decreases with increasing $t$. Furthermore, even in the absence of hadronic interactions ($t = 0.6$~fm/$c$), the $K^{*0}/K$ ratio decreases with increasing $N_{\rm part}$ and remains in good agreement with the experimental data. These observations challenge the conventional interpretation that hadronic rescattering is the dominant mechanism responsible for the suppression of the $K^{*0}/K$ ratio in central heavy-ion collisions.

%%%%%%%%%%%%%%%%%Next part is flow %%%%%%%%%%%%%%%%%%%%
\subsection{Directed and Elliptic Flow}

\begin{figure*}
\includegraphics[scale=0.75]{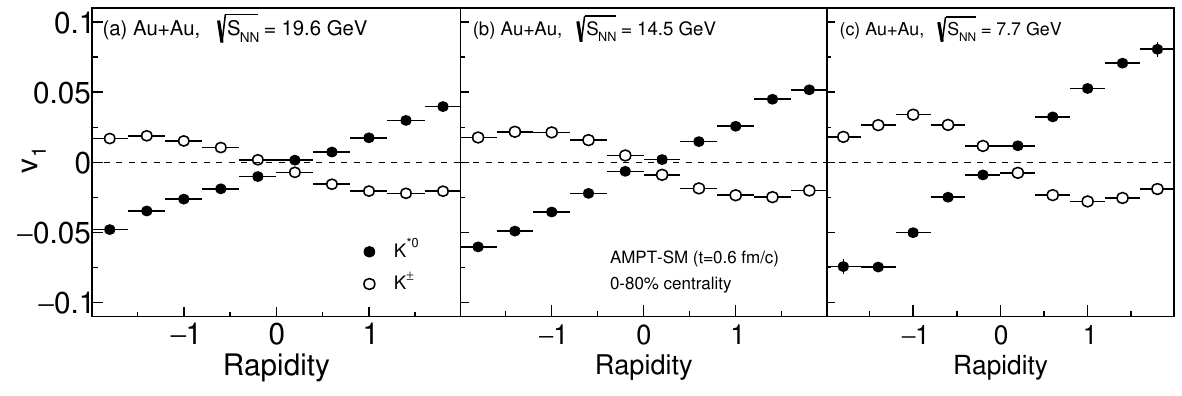}
\caption{  The directed flow ($v_1$) of $K^{*0}$ mesons and charged kaons as a function of rapidity in minimum-bias Au+Au collisions at $\sqrt{s_{NN}} = 19.6$, 14.5, and 7.7~GeV measured using  AMPT string-melting model with hadronic interactions turned off.}
\label{fig4}
\end{figure*}

\begin{figure*}
\includegraphics[scale=0.75]{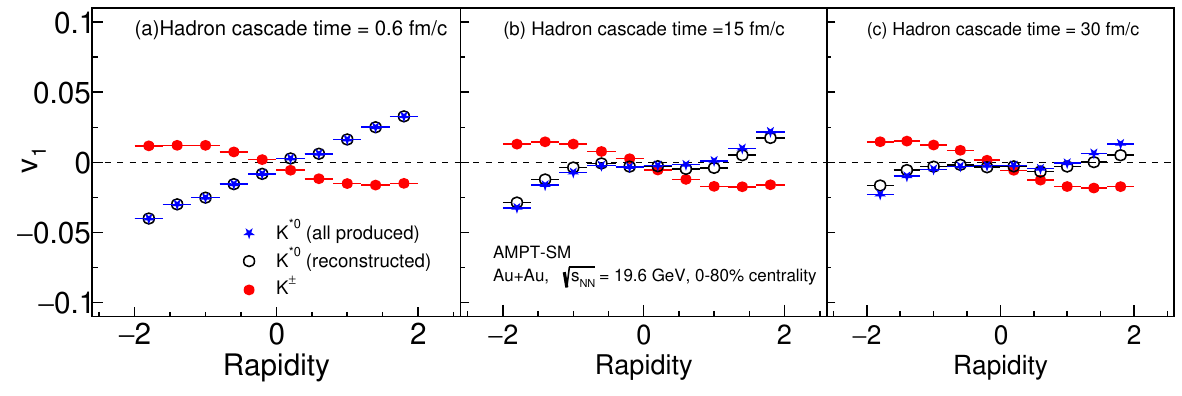}
\caption{ Rapidity dependence of the directed flow of $K^{*0}$ mesons and charged kaons, for hadronic cascade times of $t = 0.6$, 15, and 30~fm/$c$ in Au+Au collisions at $\sqrt{s_{NN}} $ = 19.6 GeV using AMPT-SM model.}
\label{fig5}
\end{figure*}

Azimuthal anisotropic flow coefficients serve as a key observable for probing the collective behaviour in heavy-ion collisions. In this framework, the initial spatial anisotropies of the collision geometry are converted into momentum-space anisotropies of the produced particles through the collective expansion of the medium~\cite{Ollitrault:1992bk, Poskanzer:1998yz}. These anisotropies are commonly characterised using a Fourier decomposition of the azimuthal particle distribution,
\begin{equation}
\frac{dN}{d\phi} \propto 1 + 2\sum_{n=1}^{\infty} v_{n}\cos n(\phi - \Psi_{n}),
\end{equation}
where $v_n$ denotes the magnitude of the $n$th-order flow coefficient and $\Psi_n$ represents the corresponding event-plane angle. Among these coefficients, the first two harmonics—directed flow ($v_1$) and elliptic flow ($v_2$)—have been studied extensively owing to their strong sensitivity to the early-time dynamics of the system. It is now well established that directed flow consists of two components: a rapidity-odd component and a rapidity-even component, the latter being independent of the reaction plane and sensitive to initial-state fluctuations. The present work focuses exclusively on the rapidity-odd component of directed flow, denoted as $v_1$.

\begin{figure*}
\includegraphics[scale=0.75]{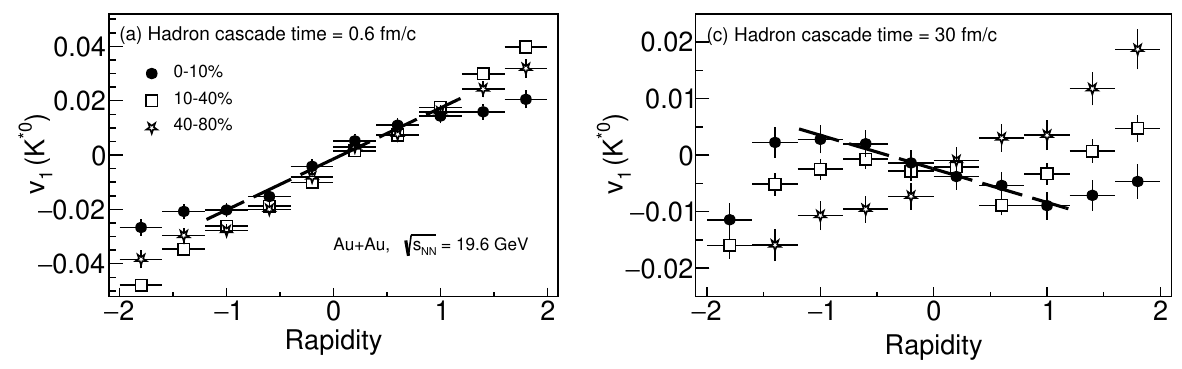}
\caption{ Centrality dependence of the directed flow of $K^{*0}$ mesons (reconstructed)  for hadronic cascade times of $t = 0.6$, and 30~fm/$c$ in Au+Au collisions at $\sqrt{s_{NN}} $ = 19.6 GeV using AMPT-SM model.  The dashed lines are fit to the $v_{1}$ using a linear function. }
\label{fig5_cent}
\end{figure*}

\begin{figure}
\includegraphics[scale=0.35]{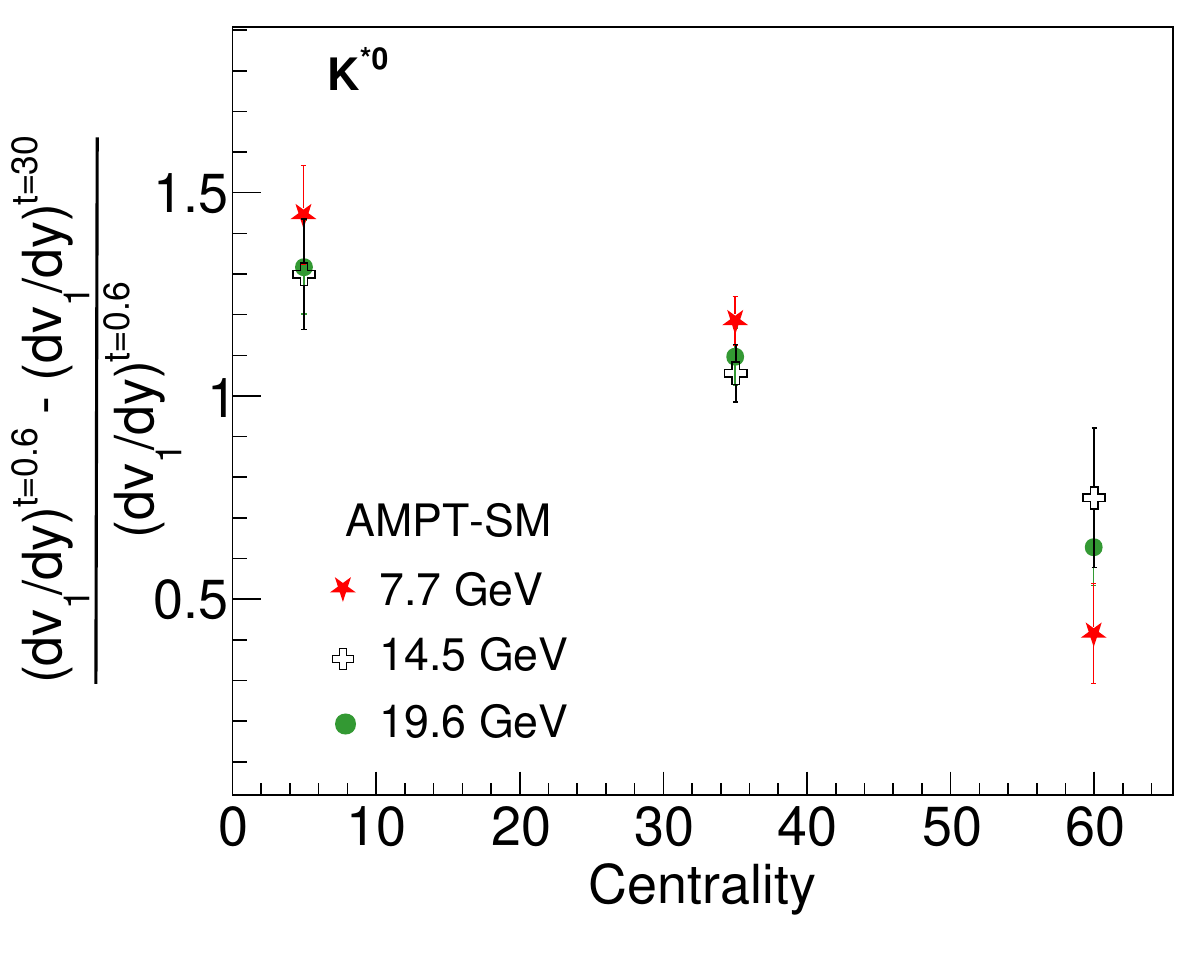}
\caption{ Centrality dependence of the slope difference of directed flow of $K^{*0}$ mesons (reconstructed)  between hadronic cascade times of $t = 0.6$, and 30~fm/$c$ in Au+Au collisions at $\sqrt{s_{NN}} $ =  7.7, 1.45 and19.6 GeV using AMPT-SM model. The observed slope difference is normalised to its value obtained at $t = 0.6$ fm/$c$.}
\label{fig5_slope}
\end{figure}

Figure~\ref{fig4} presents the directed flow of $K^{*0}$ mesons and charged kaons as a function of rapidity in minimum-bias Au+Au collisions at $\sqrt{s_{NN}} = 19.6$, 14.5, and 7.7~GeV. The results are obtained using the AMPT string-melting model with hadronic interactions turned off (i.e., $t = 0.6$~fm/$c$). It is observed that the $v_1$ of $K^{*0}$ exhibits a positive slope, in contrast to the negative slope observed for charged kaons. Furthermore, the slope of the $K^{*0}$ directed flow increases with decreasing collision energy. Similar obervation is found when $v_{1}$ is measured in 0-10\%, 10-40\% and 40-80\% centralities. In the absence of both partonic and hadron rescatterings, $v_{1}$ is found to be consistent with zero for all particles in the AMPT model. 
An opposite slope of $v_1$ between $K^{*0}$ and charged kaons has also been reported in recent preliminary measurements by the STAR experiment at RHIC~\cite{kstar_data_v1}. Since the AMPT-SM calculations with $t = 0.6$~fm/$c$ include flow generated exclusively during the partonic stage, these results indicate that the observed opposite slope between charged kaons and $K^{*0}$ resonances can arise from partonic-medium dynamics alone, even in the absence of hadronic rescattering.

Furthermore, the effect of hadronic interactions on the directed flow of $K^{*0}$ mesons and charged kaons is investigated by incorporating the hadronic cascade phase within the AMPT string-melting model. Figure~\ref{fig5}  shows the rapidity dependence of the directed flow of $K^{*0}$ mesons and charged kaons, for hadronic cascade times of $t = 0.6$, 15, and 30~fm/$c$ in 0-80\%  minimum bias Au+Au collisions at $\sqrt{s_{NN}} $ = 19.6 GeV.
When hadronic interactions are included, the numbers of produced and reconstructable $K^{*0}$ mesons differ; therefore, the directed flow ($v_1$) of produced and reconstructable $K^{*0}$ mesons is presented separately. Only a small difference is observed between the $v_1$ values of produced and reconstructable $K^{*0}$. Notably, with increasing hadronic cascade time, the magnitude of the $K^{*0}$ directed flow changes significantly, whereas the directed flow of charged kaons remains nearly unchanged. 
Centrality dependence of directed flow of $K^{*0}$ mesons (reconstructed)  as a function of rapidity is shown in Fig~\ref{fig5_cent} in Au+Au collisions at $\sqrt{s_{NN}} $ = 19.6 GeV for hadronic cascade times of $t = 0.6$,  and 30~fm/$c$.
In the absence of hadronic interactions, the $K^{*0}$ $v_{1}$ shows a weak dependence on collision centrality, and the slope $dv_{1}/dy$ remains nearly unchanged. In contrast, when hadronic interactions are included ($t = 30$~fm/$c$), a pronounced centrality dependence of $v_{1}$ is observed. With increasing centrality, the slope $dv_{1}/dy$ evolves, and changes sign in the most central (0–10\%) collisions. This behaviour is qualitatively consistent with preliminary STAR results~\cite{kstar_data_v1}, although the data exhibit an opposite trend: the slope is negative in 40–80\% centrality, and changes sign in the mid-central (10–40\%) collisions. These model results demonstrate that hadronic interactions can significantly modify, and even reverse, the directed flow of short-lived resonances such as the $K^{*0}$.Figure~\ref{fig5_slope} summarizes the difference in the slope ($dv_{1}/dy$) of $K^{*0}$ directed flow  obtained from AMPT-SM calculations with ($t = 30$ fm/$c$) and without ($t = 0.6$ fm/$c$) a hadronic cascade for Au+Au collisions at $\sqrt{s_{NN}} = 7.7$, 14.5, and 19.6 GeV. The slope ($dv_{1}/dy$) is extracted by fitting the rapidity dependence of $v_{1}$ with a linear function within $|y| < 1.0$. The observed slope difference is normalised to its value obtained at $t = 0.6$ fm/$c$. The normalised slope difference is found to be approximately independent of the collision energies considered in this study.

\begin{figure*}
\includegraphics[scale=0.75]{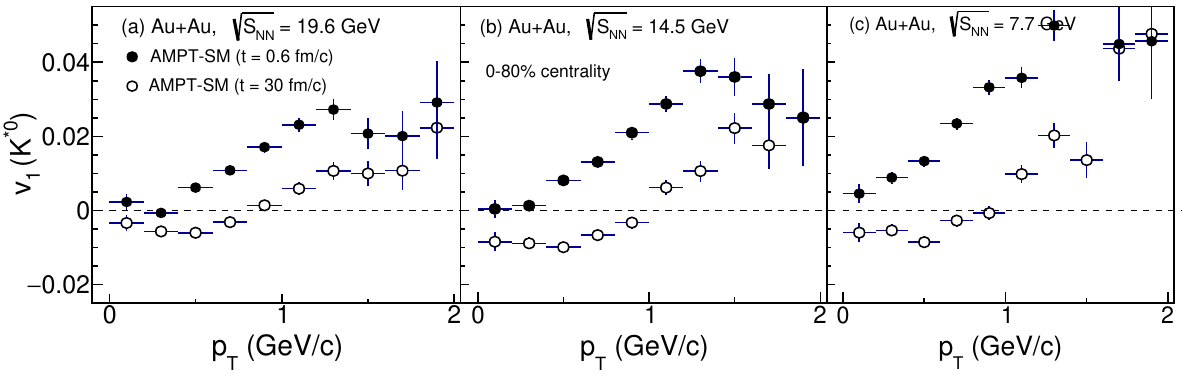}
\caption{  The transverse momentum ($p_{T}$) dependence of the directed flow of   $K^{*0}$ mesons (reconstructed)   in minimum-bias Au+Au collisions at $\sqrt{s_{NN}} = 19.6$, 14.5, and 7.7~GeV.   The results are obtained using the AMPT string-melting model with hadronic cascade times of $t = 0.6$ and 30~fm/$c$. }
\label{fig6}
\end{figure*}

\begin{figure*}
\includegraphics[scale=0.75]{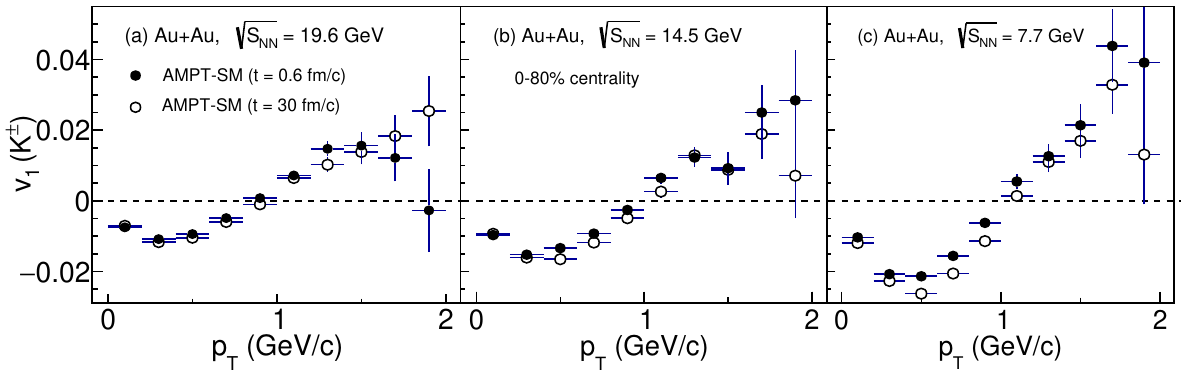}
\caption{The transverse momentum ($p_{T}$) dependence of the directed flow of $K^{\pm}$  in minimum-bias Au+Au collisions at $\sqrt{s_{NN}} = 19.6$, 14.5, and 7.7~GeV.   The results are obtained using the AMPT string-melting model with hadronic cascade times of $t = 0.6$ and 30~fm/$c$. }
\label{fig7}
\end{figure*}

Figures~\ref{fig6} and \ref{fig7} present the transverse momentum ($p_{T}$) dependence of the directed flow of $K^{*0}$ mesons and charged kaons in minimum-bias Au+Au collisions at $\sqrt{s_{NN}} = 19.6$, 14.5, and 7.7~GeV. The results are obtained using the AMPT string-melting model with hadronic cascade times of $t = 0.6$ and 30~fm/$c$ within rapidity $|y| < 1.0$. Since the odd component of $v_{1}$ by definition satisfies the symmetry $v_{1}(-y, p_{T}) = -v_{1}(y, p_{T})$, the integral of $v_{1}(y, p_{T})$ over a symmetric $y$ range yields zero. Therefore, when presenting the rapidity-integrated $v_{1}(p_{T})$, the values of $v_{1}$ at negative $y$ are multiplied by $-1$ to account for this odd symmetry. 
We observed that the directed flow of charged kaons, $v_{1}(p_{T})$, exhibits negative values at low $p_{T}$ and changes sign with increasing $p_{T}$. Moreover, the $v_{1}(p_{T})$ of charged kaons shows little dependence on the hadronic cascade time. In contrast, the $v_{1}(p_{T})$ of $K^{*0}$ mesons displays a strong sensitivity to the hadronic cascade time. In the absence of hadronic interactions, the $v_{1}(p_{T})$ of $K^{*0}$ remains positive and increases with $p_{T}$. However, when hadronic interactions are included ($t = 30$~fm/$c$), the $v_{1}$ of $K^{*0}$ at low $p_{T}$ changes sign. 

\begin{figure*}
\includegraphics[scale=0.75]{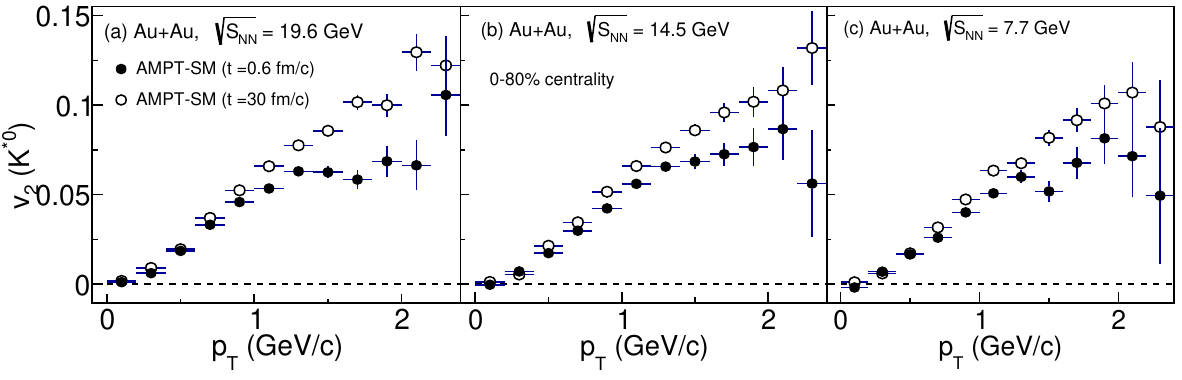}
\caption{The transverse momentum ($p_{T}$) dependence of the elliptic flow of $K^{*0}$ mesons (reconstructed)   in minimum-bias Au+Au collisions at $\sqrt{s_{NN}} = 19.6$, 14.5, and 7.7~GeV.   The results are obtained using the AMPT string-melting model with hadronic cascade times of $t = 0.6$ and 30~fm/$c$. }
\label{fig8}
\end{figure*}

%%%%%%%%%% v2%%%%%%%
Finally, the effect of hadronic rescattering on the elliptic flow of $K^{*0}$ mesons is shown in Fig.~\ref{fig8}. The transverse momentum ($p_{T}$) dependence of the elliptic flow, $v_2$, of $K^{*0}$ mesons in minimum-bias Au+Au collisions at $\sqrt{s_{NN}} = 19.6$, 14.5, and 7.7~GeV is calculated using the AMPT string-melting model with hadronic cascade times of $t = 0.6$ and 30~fm/$c$. In contrast to the directed flow, the low-$p_{T}$ elliptic flow of $K^{*0}$ mesons shows negligible sensitivity to the hadronic cascade time. Some noticeable differences, however, are observed in $v_2$ at intermediate $p_{T}$.
Overall, these results indicate that the $p_{T}$-differential directed flow, $v_1(p_{T})$, of $K^{*0}$ mesons is a more sensitive probe of hadronic rescattering effects than the corresponding elliptic flow, $v_2(p_{T})$.

\section{Summary}
%%%%%%%%%%%%%%
We have investigated the production and collective flow properties of $K^{*0}$ resonances in Au+Au collisions at $\sqrt{s_{NN}} = 19.6$, 14.5, and 7.7~GeV using the AMPT model. Both the string-melting (AMPT-SM) and default (AMPT-Def) configurations were employed to study the effects of hadronic rescattering on the $K^{*0}$ yield, $\langle p_{T} \rangle$ , the $K^{*0}/K$ ratio, and anisotropic flow observables.
Our results show that hadronic rescattering leads to a significant suppression of the reconstructable $K^{*0}$ yield, particularly at low transverse momentum, with the suppression increasing for longer hadronic cascade times. While this effect is stronger in the AMPT-Def model, the centrality dependence of the $K^{*0}/K$ ratio is better reproduced by the AMPT-SM model, whereas the AMPT-Def model overpredicts the experimental measurements. Notably, within the AMPT-SM framework, the $K^{*0}/K$ ratio is found to be nearly independent of the hadronic cascade time, and the experimental measurement is already well described even in the absence of hadronic interactions. This observation challenges the conventional interpretation that hadronic rescattering alone drives the suppression of the $K^{*0}/K$ ratio in central collisions. In addition, we find that  the average transverse momentum, $\langle p_{T} \rangle$, of  $K^{*0}$ is sensitive to the lifetime of hadronic phase, increasing significantly with the extension of the hadronic phase lifetime.

We further show that the directed flow ($v_1$) of $K^{*0}$ mesons is highly sensitive to both partonic dynamics and hadronic rescattering. In particular, the opposite slope of $v_1$ between $K^{*0}$ and charged kaons can arise even in the absence of hadronic interaction, while hadronic rescattering significantly modifies the magnitude of $K^{*0}$ $v_1$. In contrast, the elliptic flow ($v_2$) exhibits only a weak sensitivity to hadronic interactions. These findings establish $K^{*0}$ directed flow as a particularly sensitive probe of the late-stage hadronic medium in heavy-ion collisions.

% BibTeX users, please use
% \bibliographystyle{}
% \bibliography{}
%
% Non-BibTeX users please use
%\begin{thebibliography}{}

\end{document}